\documentstyle[eqsecnum,aps,prd,epsf]{revtex}
\date{9 September 1998}
\newcommand\lsim{\mathrel{\rlap{\lower4pt\hbox{\hskip1pt$\sim$}}
    \raise1pt\hbox{$<$}}}
\newcommand\gsim{\mathrel{\rlap{\lower4pt\hbox{\hskip1pt$\sim$}}
    \raise1pt\hbox{$>$}}}

\tighten
\begin{document}
\title{Cosmological Consequences of String-forming Open Inflation Models}

\author{P. P. Avelino${}^{1}$\thanks{
Electronic address: pedro\,@\,astro.up.pt},
R. R. Caldwell${}^{2}$\thanks{
Electronic address: caldwell\,@\,dept.physics.upenn.edu},
and C. J. A. P. Martins${}^{3}$\thanks{Also at C. A. U. P.,
Rua do Campo Alegre 823, 4150 Porto, Portugal.
Electronic address: C.J.A.P.Martins\,@\,damtp.cam.ac.uk}}

\address{${}^1$ Centro de Astrof\'{\i}sica, Universidade do Porto\\
Rua das Estrelas s/n, 4150 Porto, Portugal}

\address{${}^2$ Department of Physics and Astronomy, University of
Pennsylvania\\
Philadelphia, PA 19104}

\address{${}^3$ Department of Applied Mathematics and Theoretical Physics\\
University of Cambridge, Silver Street, Cambridge CB3 9EW, U.K.}

\maketitle
\begin{abstract}
We present a study of open inflation cosmological scenarios in which
cosmic strings form betwen the two inflationary epochs.
It is shown that in these models strings are stretched outside the horizon
due to the inflationary expansion but must necessarily re-enter
the horizon before the epoch of equal matter and radiation densities. We
determine the power spectrum of cold dark matter perturbations in these hybrid
models, finding good agreement with observations for values of
$\Gamma=\Omega_0h\sim0.3$ and comparable contributions from the active and
passive sources to the CMB. Finally, we briefly discuss other cosmological consequences
of these models.

\end{abstract}

\section{Introduction}
\label{sint}

Over the last two decades, two paradigms have emerged as possible
explanations for the origin of the cosmological structures seen today.
Inflationary
models \cite{inf}, originally developed with the aim of solving the horizon
and flatness problems, were soon seen to generate a spectrum of
quantum-mechanical fluctuations with an approximately scale-invariant
(Harrison-Zel'dovich) form. On the other hand, cosmic strings \cite{vsh}
are topological defects which may have formed at cosmological phase
transitions, and since their late-time evolution is
scale-invariant they to can produce a scale-invariant spectrum of
perturbations.

These two paradigms are usually considered to be incompatible with each
other, the argument being roughly that any strings formed before or during
inflation will be stretched to unobservable scales, and any strings formed
afterwards will be too light to have any cosmological relevance (at least,
from the point of view of structure formation).

Recently, however, theoretical cosmologists have realized (aided,
undoubtedly, by their observational counterparts \cite{obs}) that inflation
does not necessarily produce a critical-density universe, and in fact
inflationary models can now be built that produce any value of $\Omega_0$
(subject to some model-dependent amount of fine-tuning) \cite{bgt,alm}. In
particular, there has been recent interest in models where an open universe
is produced as a result of two epochs of inflation, and it has been
suggested that in such models cosmic strings can form between the these two
epochs \cite{tdi}.

Here we present a study of the cosmological consequences of these models.
Since the second inflationary period is rather short, it does not
necessarily follow that they will be unobservable by the present day. In
fact, it will be shown that in these models strings must necessarily
re-enter the horizon before the epoch of equal matter and radiation densities.
This means that they can still play a crucial cosmological role, namely in the
formation of structures in the universe. Note that these will therefore be
hybrid structure formation models, with perturbations being generated by
both active and passive sources.

We will start by reviewing the key concepts behind open inflation in
section \ref{sinf}. Following this we discuss cosmic string evolution in
these models (section \ref{sstr}) and calculate the corresponding
total CDM power spectrum (section \ref{slss}). Finally, we briefly discuss
other cosmological consequences of these models and present our conclusions.
Throughout this paper we have chosen units such that the speed of light is
unity, $c = 1$, and adopted the convention  $H_o \equiv 100\, h \,{\rm
km}\, {\rm sec}^{-1}\,{\rm Mpc}^{-1}$.

\section{Basics of Open Inflation}
\label{sinf}

Open inflation models can be roughly described as a period of old inflation
followed by a period of new inflation \cite{bgt,alm}. One starts with a
scalar field trapped in false vacuum. The potential barrier must be high
enough for the corresponding decay rate to be exponentially suppressed.
This is required for the universe to expand enough for bubble nucleation to
occur in a smooth de Sitter spacetime background. Additionally, this
ensures that the probability of collision between two bubbles is small
enough for the universe to survive up to the present day.

In these circumstances, the interior of the bubble will be a homogeneous
and isotropic universe (thus solving the horizon problem) with negative
spatial curvature. Then the inflaton starts rolling down the potential,
preserving homogeneity and isotropy, and as it approaches the true minimum
its oscillations will reheat the universe and leave it in the usual
radiation-doinated epoch. The value of the present density of the universe
will be determined by the duration of this second inflationary period,
which is in turn determined by the initial value of the inflaton.

Note that the two periods of inflation may \cite{bgt} or may not \cite{alm}
be driven by the same scalar field. In the latter case the first
inflationary epoch will be due to a `heavy' inflaton with a steep potential
barrier, while in the second epoch a `light' inflaton rolls in a
nearly-flat direction orthogonal to that of the original tunnelling.
Also, in the latter case we will typically have an ensemble of very large
(but finite) 'inflationary islands' \cite{gbgm} rather than an infinite
open universe. While
the specific particle-physics details of the two classes of models will of
course be different, their basic cosmological features (to be summarized
below) are nevertheless very similar. Hence, for simplicity we will
be concentrating on the
model by Bucher {\em et al.} \cite{bgt} in the remainder of this paper.

Now, the metric inside the nucleated bubble has the usual
Friedmann-Robertson-Walker (FRW) form
\begin{equation}
ds^2=-dt^2+a^2(t) \left[d\zeta^2+\sinh^2(\zeta)d\Omega_2^2\right]\, ,
\label{metric}
\end{equation}
where $d\Omega_2^2=d\theta^2+\sin^2\theta d\phi^2$.
Then the corresponding Friedmann and inflaton equations are
\begin{equation}
H^2 =\left(\frac{\dot{a}}{a}\right)^2=\frac{1}{a^2}+\frac{8\pi}{3m_{Pl}^2}
\left(\frac{1}{2}\dot{\phi}^2 + V(\phi)\right)\, ,
\label{friedmann}
\end{equation}
\begin{equation}
\ddot{\phi}+3H\dot{\phi}= -\frac{\partial V}{\partial\phi}\, .
\label{inflaton}
\end{equation}
We will take the potential to be of the generic form
\begin{equation}
V(\phi)=\frac{\lambda\phi^n}{nm_{Pl}^{n-4}}\, .
\label{potential}
\end{equation}
but for definiteness we will generally use $n=2$ in the following sections.
It can be shown \cite{bgt} that for given choices of $n$, the temperature
of reheating (denoted $T_r$) and the cosmological parameters $\Omega_0$ and
$h$, the appropriate open inflationary model has an initial value of the
inflaton given by
\begin{equation}
f^2\equiv\left(\frac{\phi}{m_{Pl}}\right)^2_i=\frac{n}{8\pi}\left[49.48+
\ln{\left(\frac{T_r(GeV)^2}{h^2(1-\Omega_0)}\right)}\right]\, .
\label{startphi}
\end{equation}

It is fairly straightforward to see that there are two evolution regimes
for the above equations. At early times, spatial curvature will be
dominant, so we have
$a\propto t$ and $\Omega\simeq0$. Then the second epoch of inflation will
set in, and the scale factor will approximately evolve as
$a\propto\exp{(t)}$, with $\Omega$ being driven to a value very close to
unity. The transition between these two regimes happens at a time $t_t$
given by
\begin{equation}
\frac{t_r}{t_t}=1+\frac{4\pi f^2}{n}=N\, ,
\label{inftwo}
\end{equation}
where $t_r$ is the epoch corresponding to the reheating temperature $T_r$.
Note that this ratio is also the number of e-foldings N in the model.
As the inflaton approaches its true minimum, reheating takes place. We will
assume that this happens instantaneously, and therefore the universe
rapidly becomes radiation-dominated, with the scale factor growing as
$a\propto t^{1/2}$. By numerically studying equations (\ref{friedmann}) and
(\ref{inflaton}), one can show that the following analytic expression
provides a reasonably accurate description of the evolution of the scale
factor
\begin{equation}
a(t) = \cases{
\frac{t}{t_{Pl}} & for $t<t_t$\cr
\frac{t_t}{t_{Pl}} {\rm exp}\left(\frac{t}{t_t}-1\right) & for $t_t<t<t_r$\cr
\frac{t_t}{t_{Pl}} {\rm
exp}\left(\frac{t_r}{t_t}-1\right)\left(1+2\frac{t-t_r}{t_t}\right)^{1/2} &
for $t_r<t<t_{eq}$
}\, . \label{scalefactor}
\end{equation}
The Hubble length has the form
\begin{equation}
H(t)^{-1}= \cases{
t & for $t<t_t$\cr t_t & for $t_t<t<t_r$\cr
2(t-t_r)+t_t & for $t_r<t<t_{eq}$
}\, . \label{hubblec}
\end{equation}
These will therefore be used in the remainder of the paper.

\section{String Formation and Evolution}
\label{sstr}

As discussed by Vilenkin \cite{tdi}, there are a number of reasonably
natural mechanisms that could lead to string formation between the two
inflationary epochs. Two simple exemples are the cases where a complex
scalar field is coupled to the field responsible for the first inflationary
epoch or to spatial curvature. Again we will not be discussing the
individual details of each possible model but will instead focus on their
generic cosmological properties.

The evolution of the cosmic string network will be described in the context
of the quantitative analytic model introduced by Martins and Shellard
\cite{ms2} and extended to general FRW models in by the present authors
\cite{acm}. Here we will briefly recall the essential features of this
approach; the reader is referred to the above papers for a more
detailed discussion (see also \cite{phd}).

The basic idea is to describe the string network by a small number
of macroscopic (or `averaged') quantities whose evolution equations
are derived from the microscopic string equations of motion. In the
simplest case, there are only two such quantities, the energy of a piece of
string,
\begin{equation}
E=\mu a(\tau)\int\epsilon d\sigma\, ,
\label{defne}
\end{equation}
($\epsilon$ being the coordinate energy per unit length on the string
worldsheet), and the string root-mean squared (RMS) velocity, defined by
\begin{equation}
v^2=\frac{\int{\dot{\bf x}}^2\epsilon d\sigma}{\int\epsilon d\sigma}
\, . \label{defnv}
\end{equation}
In passing, we point out that this kind of approach has been extended to
more complicated models, which are suitable for describing cosmic strings
containing significant small-scale strucutre (`wiggles') \cite{phd,msw} or
superconducting currents \cite{phd,mss}. In the present case, however, the
inflationary epoch will make any such effects less important than in more
standard scenarios, so we will not explicitly consider them (however, see
section \ref{slss}).

Distinguishing between long (or `infinite') strings
and loops, and knowing that the former should be Brownian
we can define the long-string correlation length as
$\rho_{\infty}\equiv\mu/L^2$. A phenomenological term must then be included
describing the interchange of energy between long strings and loops---a
`loop chopping efficiency', denoted ${\tilde c}$ and expected to be
slightly smaller than unity. One can then derive the evolution equation for
the
correlation length $L$, which has the form \cite{ms2,acm}
\begin{equation}
\frac{dL}{dt}=HL\left(1+\frac{v^2}{\left[1-(1-\Omega)(HL)^2\right]^2}\right)+
\frac{{\tilde c}}{2}v \, . \label{evoll}
\end{equation}
Note that we do not include frictional forces due to particle scattering on
strings \cite{ms2}, since for cosmic strings forming at or around the GUT
scale these are subdominant after reheating.

One can also derive an evolution equation for the long string
velocity with only a little more than Newton's second law \cite{ms2,acm}
\begin{equation}
\frac{dv}{dt}=\left[\left[1-(1-\Omega)(HL)^2\right]^2-v^2\right]\frac{{\tilde k}
}{L}-2Hv\left(1-\frac{v^2}{\left[1-(1-\Omega)(HL)^2\right]^2}\right) \, ;
\label{evolv}
\end{equation}
here ${\tilde k}$ is another phenomenological parameter of order unity
which  specifies the relative importance of curvature and momentum in
determining the dynamics of the strings \cite{ms2}.
These quantities are sufficient to quantitatively
describe the large-scale properties a cosmic string network.

A simple analysis of the evolution equations (\ref{evoll})--(\ref{evolv})
reveals the basic features of the evolution of a cosmic string network in
these circumstances. Since we are mainly interessed in these strings as
possible seeds for structure formation, we will assume later on
that they are formed
at an epoch such that they have a mass per unit length $G\mu\sim10^{-6}$. It
is fairly straightforward to generalize these results to other energy
scales---see for exemple \cite{ms2,phd}. Before the second
inflationary period starts we have
$\Omega\simeq0$ and an approximate solution (meaning that it would be exact if
$\Omega$ was exacly zero) is \cite{cmopen}
\begin{equation}
L(t)\propto t \left(\ln{t}\right)^{1/2}\, ,
\label{soll1}
\end{equation}
\begin{equation}
v(t)\propto \left(\ln{t}\right)^{-1/2}\, .
\label{solv1}
\end{equation}
Hence, at the onset of the second inflationary epoch, strings will be
fairly straight (having a correlation length close to the horizon) and
non-relativistic. It is intuitively obvious that during inflation the strings
will be conformally stretched, and indeed the solution of
(\ref{evoll})--(\ref{evolv}) in this case is
\begin{equation}
L(t)\propto \exp{\left(\frac{t}{t_t}\right)}\, ,
\label{soll2}
\end{equation}
\begin{equation}
v(t)\propto \exp{\left(-\frac{t}{t_t}\right)}\, .
\label{solv2}
\end{equation}
Note that it is only during the second period of inflation (not before) that
the strings are stretched outside the horizon. This fact is crucial for
what follows.

It is easy to verify using the expressions for the evolution of the scale
factor and Hubble parameter that if a given scale leaves the Hubble sphere at
the time
$t_{out}<t_r$ and re-enters it at the time $t_{in} \gg t_r$
then
\begin{equation}
{a_{in} \over a_r} \times {a_r \over a_{out}}
\sim {\left({{2 t_{in}} \over t_t}\right)}^{1/2}
\times {a_r \over a_{out}} \sim {{2 t_{in}} \over t_t}\, , \label{demo1}
\end{equation}
which implies that
\begin{equation}
{a_{in} \over a_r} \sim {a_r \over a_{out}}\, . \label{demo2}
\end{equation}
We also know that during inflation the importance of the curvature term
decreases as
\begin{equation}
\Omega-1={k \over {a^2 V}} \propto a^{-2}\, , \label{demo3}
\end{equation}
while during the radiation era we have
\begin{equation}
\Omega-1={k \over {a^2 \rho_{rad}}} \propto a^{2}\, . \label{demo4}
\end{equation}
This implies that the value of $\Omega$ when the strings leave the horizon
must be approximatelly the same to the value of the $\Omega$ when the strings
re-enter the horizon. Consequently, since they leave during the second
inflationary period, when the density is very close to critical,
we can expect them to re-enter not much later than the epoch of equal
radiation and matter densities. This can be confirmed as follows.

From the above discussion, we know that at the start of the radiation
era the strings will have
negligible velocity (as well as small-scale structure) and will continue to
be conformally stretched, so that our averaged quantities now behave as
\begin{equation}
L(t)\propto t^{1/2}\, ,
\label{soll3}
\end{equation}
\begin{equation}
v(t)\propto t^{1/2}\, .
\label{solv3}
\end{equation}
Notice the different behaviour of the string RMS velocity in the present
case compared to the usual `stretching regime' (where it behaves like
$v\propto t$---see \cite{ms2,phd}). This is because, as we pointed out
above, in the present case the effect of the frictional force due to
particle scattering on strings will be negligible. Obviously these
solutions can be confirmed numerically (see below).

Now, the crucial observation is that this slow growth of the correlation
length $L$ relative to $H^{-1}$ (which is proportional to time in the
radiation era) means that strings can indeed get back inside the horizon
in time to play a significant role in the formation of structures in the
universe. As velocity grows the loop formation term will
eventually become important, and
the network will `switch' to the linear scaling regime. In principle, one
would require that strings should be back inside the horizon at the epoch
of equal matter and radiation densities, $t_{eq}$. However, we note that
it is also
clear that some time will be necessary for the network to reach the usual
linear scaling regime \cite{ms2} once it is back inside the horizon. Hence
one sees that this class of models will provide a rather natural way of
generating `deviations from scaling' near $t_{eq}$, which have been claimed
\cite{abr} to be necessary to reconcile the cosmic string scenario with
observation.

For the moment, we can estimate how much inflation can be tolerated by a
cosmic string network that is requred to re-enter the horizon before
$t_{eq}$. We require $H(t_{eq})^{-1} > L(t_{eq})$ to determine
the amount of inflation which may occur and define
$N$ to be the number of e-foldings in the second inflationary period. We
assume for simplicity that the correlation length changes instantaneously
from the exponential to the power-law growth laws. Also for simplicity, we
temporarily consider the case where the second inflationary period starts
at the moment when the strings form.

\begin{figure}
\vbox{\centerline{
\epsfxsize=0.6\hsize\epsfbox{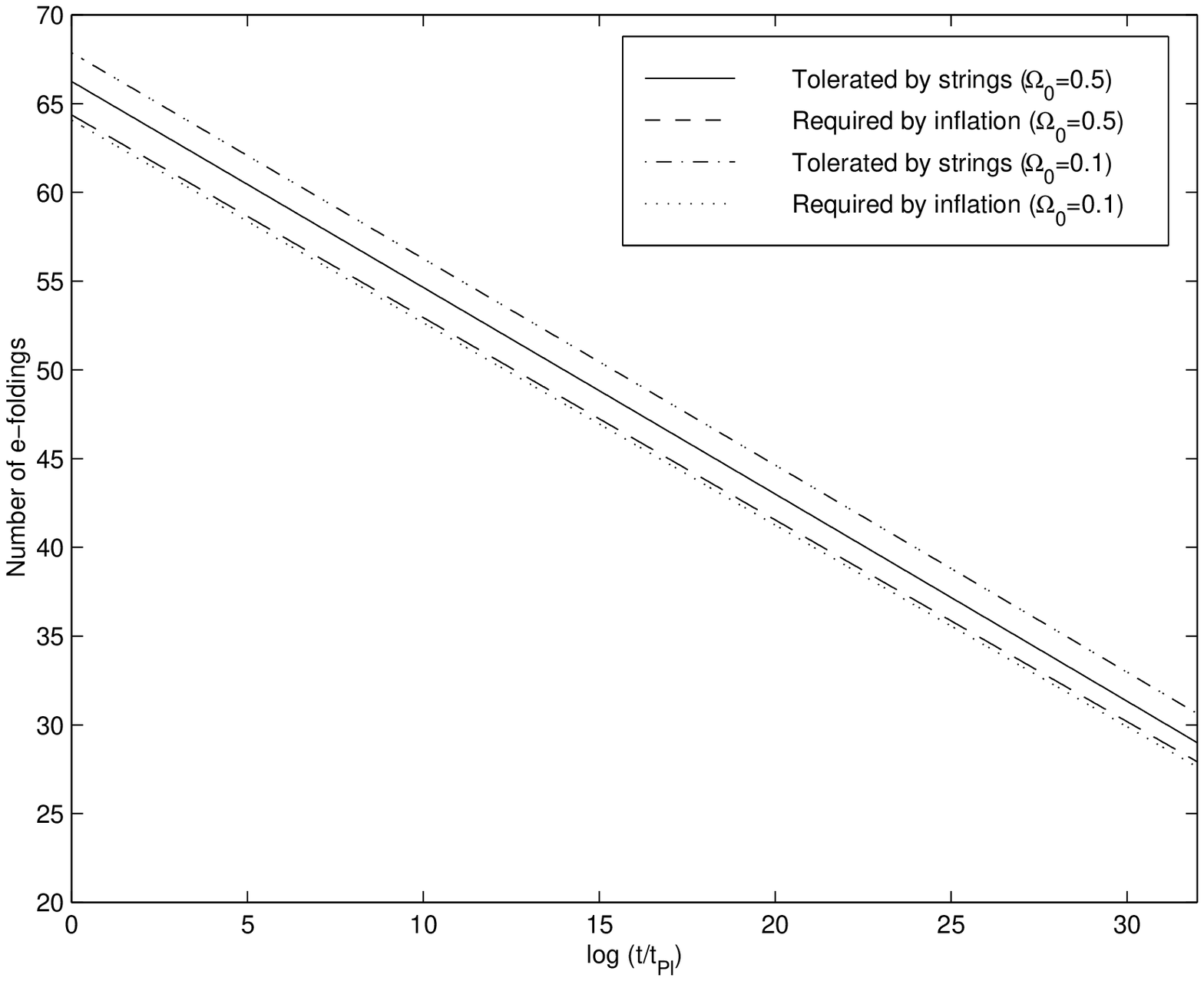}}
\vskip.4in}
\caption{The maximum number of efoldings tolerated by a string network
(formed at time $t$ shown in the horizontal axis) that
is required to be inside the horizon at $t_{eq}$
and the number of efoldings in an open inflationary model such
that the second inflationary period starts when the strings form and the
present density is $0.5$ or $0.1$.}
\label{figefold}
\end{figure}

\begin{figure}
\vbox{\centerline{
\epsfxsize=0.6\hsize\epsfbox{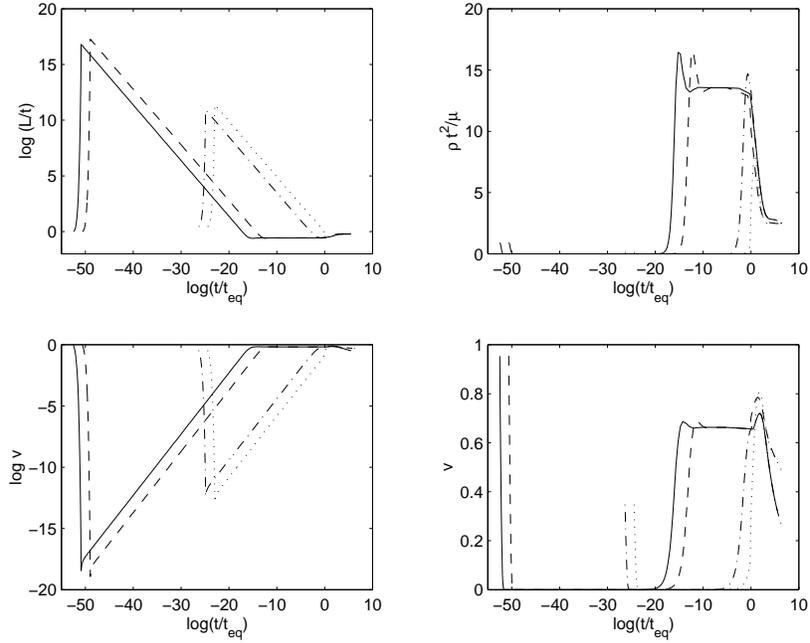}}
\vskip.4in}
\caption{Evolution of GUT-scale string networks in open inflation models, for
the following sets of initial conditions: $T_r=10^{15}\,GeV, \Omega_0=0.1$
(solid lines), $T_r=10^{15}\,GeV, \Omega_0=0.9$ (dashed), $T_r=10^2\,GeV,
\Omega_0=0.1$ (dash-dotted) and $T_r=10^2\,GeV, \Omega_0=0.9$ (dotted).}
\label{figsamp}
\end{figure}

It is then a simple matter to show that the maximum number of e-foldings
$N_{str}$ that
a string network can tolerate in these circumstances is given by the
solution to
\begin{equation}
N_{str}=\frac{1}{2}\left[\ln\left(4\frac{t_{eq}}{t_t}\right)+\ln\left(1
+N_{str}\right)\right]\, .
\label{valnstr}
\end{equation}
On the other hand, the number of e-foldings $N_{inf}$ required for an open
inflation model whose second period of inflation starts when the strings
form to produce a present universe with a density $\Omega_0$ is
\begin{equation}
N_{inf}=\frac{1}{2}\left[49.48+\ln\left(\frac{0.301m_{Pl}^2}
{h^2(1-\Omega_0)}\frac{t_{Pl}}{t_t}\right)-\ln\left(1 +N_{inf}\right)\right]\, .
\label{valninf}
\end{equation}
These two quantities are plotted (for $h=0.7$ and two different values of
$\Omega_0$) in figure \ref{figefold}.
This confirms that strings will necessarily be back inside the horizon at
$t_{eq}$ . Depending on initial conditions, they can actually return many
orders of magnitude in time before that.
Note that even though we assumed (for simplicity)
that strings were formed at $t=t_t$, our result is completely general,
since even if formed before strings
can not leave the horizon until $t_t$. From this plot we can also see that
$N_{str}$ is much more sensitive to the cosmological parameters than
$N_{inf}$ (due to its dependence on $t_{eq}$). Furthermore, since
$t_{eq}/t_{Pl}\propto\Omega_0^{-2}$, the smaller $\Omega_0$, the more
comfortably strings will get back inside. In other words, the 'extreme'
case where strings just manage to get back inside the horizon at $t_{eq}$
happens for
a critical density that is just below unity and for a reheat temperature
that is as low as possible. This can be confirmed in figure
\ref{figsamp}, where we have summarized the cosmological evolution of some
sample string networks.

\section{Large Scale Structure}
\label{slss}

We assume that during the second period of inflation the following
slow-roll conditions are satisfied:
\begin{equation}
{\dot \phi} \sim -{1 \over {3 H}} V',
\label{inf0}
\end{equation}
\begin{equation}
{\epsilon} \equiv {{m^2_{\rm pl}} \over {16 \pi}}
{\left( {{V'} \over V} \right)}^2 \ll 1,
\label{inf1}
\end{equation}
and
\begin{equation}
{\eta} \equiv {{m^2_{\rm pl}} \over {8 \pi}}
{\left( {{V''} \over V} \right)} \ll 1.
\label{inf2}
\end{equation}
In this case the condition ${\dot \phi}^2 \ll V$ is also verified
and then
\begin{equation}
{H^2} \sim {{8 \pi V} \over {3 m^2_{\rm pl}}}.
\label{inf3}
\end{equation}
As we have seen before the initial value of the inflaton field can be
chosen in such a way as to allow $\Omega_0$ to have any 'desired'
value below unit. Here, we start by reviewing the
results for the spectrum of adiabatic fluctuations produced during
the second stage of inflation subject to the three slow-roll conditions
listed above
in the case with $\Omega_0 \sim 1$. We will then discuss how the power
spectrum
can be rescaled to allow for open models taking into account that
scales $\lsim 100 h^{-1} {\rm Mpc}$ come inside
the horizon when the curvature is still dynamically
unimportant. For the sake of simplicity we will also take $h=1$ for the
time being.

The power spectrum of the density contrast
measured today of the adiabatic fluctuations produced during inflation in
a flat universe with zero cosmological constant can be written as
\begin{equation}
S(k)=4 \pi^3 P(k) = {\left(k \over {a_0 H_0} \right)}^4 T^2(k) A_S^2(k).
\label{infp}
\end{equation}
where $k$ is given in units of ${\rm Mpc}^{-1}$.
The quantity $A_S^2(k)$ specifies the initial spectrum and can be
calculated
precisely in
terms of the inflationary potential \cite{lyth} as
\begin{equation}
A_S^2(k) = {{32 V } \over {75 m^4_{\rm pl}}} \epsilon^{-1}\propto\lambda
\left(\frac{\phi}{m_{Pl}}\right)^{n+2}.
\label{infd}
\end{equation}
The transfer function $T(k)$ for the CDM models considered here is accurately
given by Bardeen et al. \cite{bbks} as
\begin{equation}
T(k)={\ln (1+2.34 k) \over {2.34 k}} \times
[1+3.89k+(16.1k)^2+(5.46k)^4+(6.71k)^4]^{-1/4}.
\label{inft}
\end{equation}

Having discussed the results in the simpler case with $\Omega_0=1$ we are
now in
a position to generalize these results for open universes.
To do this we take into account that perturbations on the scales of
interest to us, that is $\lsim 100 h^{-1} {\rm Mpc}$,
were generated and re-enter the horizon when $\Omega$ is still very close to
unity for any observationally reasonable choice of $\Omega_0$. Having
established this, it is straightforward to rescale the
spectrum for a flat universe with a zero cosmological constant in the
following way :
\begin{equation}
  \label{Sopen}
  S(k,h,\Omega_{0}) =
  S(k,1,1) \cdot
  \Omega_{0}^2 {h}^4 \cdot
  g^2(\Omega_{0}),
\end{equation}
where $k$ is in units of $\Omega_{0}h^2 \, {\rm Mpc}^{-1}$,
and $g(\Omega_{0})$ is given by \cite{cpt}
\begin{equation}
  \label{gomega}
  g(\Omega_{0})=\frac{5\Omega_{0}}
  {2\left[\Omega_{0}^{4/7}+
      1+\Omega_{0}/2\right]}.
\end{equation}
The factor $g(\Omega_{0})$ gives the supression of growth of
density perturbations in an open universe relative to that of a
flat universe with zero cosmological constant.
Let us define the horizon crossing amplitude $\delta_H(k)$ as
$$
\delta_H(k)=A_S(a_0 H_0) \times {{g(\Omega_0)} \over \Omega_0}
$$
This quantity is constrained by the four-year COBE data to be:
$$
\delta_H(\Omega_0)=1.95 \Omega_0^{-0.35-0.19 \ln \Omega_0}
$$
if only inflationary perturbations are responsible by cosmic structure.
This fit works to better than $2 \%$ for $0.2 \le \Omega_0 <1$ and the
statistical uncertainty is $7 \%$ \cite{blw}. The above normalization was obtained 
for an open model with a flat-space scale invariant spectrum instead of the 
open-bubble inflation model spectrum. However, for 
$0.3 \le \Omega_0 <1$ these lead 
to different normalization for $\delta_H$ only by a factor smaller 
than $10 \%$ \cite{grs}.

Having characterized the spectrum of adiabatic fluctuations produced during the
second stage of inflation we now proceed to study the
power spectrum generated by the cosmic string network. Again we
start by studying the simpler case of a $\Omega_0\sim1$ universe and then
generalize
our results for open models as we did for the
inflationary perturbations. We use the semi-analytic model of Albrecht and
Stebbins \cite{as} to estimate the power spectrum  of density perturbations
induced by
cosmic strings. This is given by
\begin{eqnarray}
&& P(k) = 16 \pi^2 (1 + z_{eq})^2 \mu^2 \int_{\tau_i}^\infty
|T(k;\tau')|^2 {\cal F}(k \xi/a) d\tau' \cr\cr
&& {\cal F}(k \xi/a) = {2 \over \pi^2}
{{ \beta^2 \Sigma}} {\chi^2 \over \xi^2}
\Big(1 + 2(k \chi/a)^2 \Big)^{-1} .
\label{powerspectrum}
\end{eqnarray}
In these equations, $a$ is the scale factor which evolves smoothly from
radiation- to dust-dominated expansion,  $\tau_i$ is the conformal time at
which the string network formed, and $T(k,\tau')$ is the transfer function for
the evolution of the causally-compensated perturbations. The parameters used
in the Albrecht-Stebbins estimate of the cosmic string
power spectrum are given by
\begin{equation}
\xi \equiv (\rho_L / \mu)^{-1/2}, \qquad
\beta \equiv \langle v^2 \rangle^{1/2}, \qquad
\Sigma \equiv
{\mu_r \over \mu}\gamma_b \beta_b + {1 \over 2 \gamma_b \beta_b}
\Big({\mu_r^2 - \mu^2 \over \mu \mu_r}\Big)
\end{equation}
where $\chi$ is the curvature scale of wakes, $\beta_b$ is the macroscopic bulk
velocity of string, $\gamma_b = (1 - \beta_b^2)^{-1/2}$, and $\mu_r$ is the
renormalized   mass-per-unit-length, which reflects the accumulation of small
scale structure on the string.
Here we assume that $\chi=2 \xi$,
$\beta_b= \beta / \mu_r$, $\mu_r = {(1-\beta^2)}^{-1/2}$.
We note that this semi-analytical model was shown to provide reasonably
accurate results when
compared with numerical calculations using high-resolution string network
simulations \cite{aswa,nsim}.

We generalize our results for open universes simply by rescaling the
spectrum $\Omega_0=1$ universe in a similar way to what we did
for the inflationary perturbations :
\begin{equation}
  \label{Sopen2}
  S(k,h,\Omega_{0}) =
  S(k,1,1) \cdot
  \Omega_{0}^2 {h}^4 \cdot f^2(\Omega_{0}) \cdot
  g^2(\Omega_{0}),
\end{equation}
where $k$ is in units of $\Omega_{0}h^2 \, {\rm Mpc}^{-1}$. The new factor
$f(\Omega_{0})=\Omega_{0}^{-0.3}$ reflects the dependence of
the COBE normalisation of $G\mu$ on $\Omega_{0}$,
which changes upwards as we decrease the matter density
in an open universe \cite{acm}.

We can finally give the total power by taking into account that the cosmic
string and inflationary power spectra of density fluctuations
and CMB anisotropies are uncorrelated.
In this case the total power is given by
$$
S_{\rm total}=\alpha S_{\rm inf} + (1-\alpha) S_{\rm str}.
$$
where $S_{\rm inf}$ and $S_{\rm str}$ represent respectively the
COBE normalized inflationary and string power spectrum and $\alpha$ represents
the fraction of the power which is due to adiabatic perturbations produce
during inflation.

\begin{figure}
\vbox{\centerline{
\epsfxsize=1.0\hsize\epsfbox{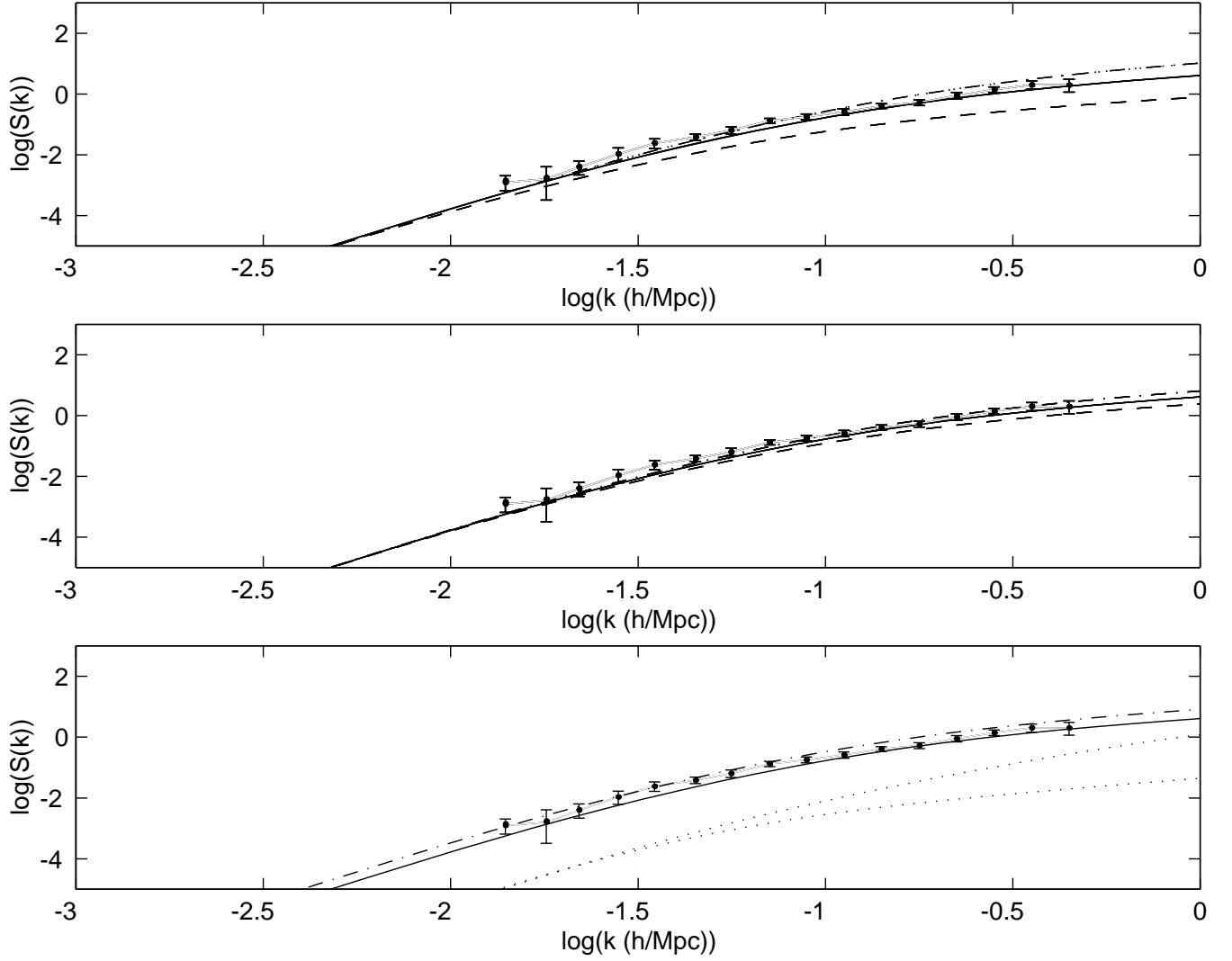}}
\vskip.4in}
\caption{Comparing the CDM hybrid (strings plus inflation) power spectrum
in open inflationary models, for several choices of $\Omega_0$, $h$
and $T_r$. In the top panel, the power spectra for $h = 0.65$, $\alpha=0.5$ and
$\Omega_0 = 0.7,\,0.5,\,0.3$ are shown (dot-dashed, solid and dashed lines respectively). 
In the middle pannel we plot the power spectra for $\Omega_0 = 0.5$, $\alpha=0.5$ and 
$h = 0.75,\,0.65,\,0.55$ (dot-dashed, solid and dashed lines respectively). The lower pannel 
represents the power spectra for $\Omega_0=0.5$, $h=0.65$ and 
$\alpha=1,\,0.5,0.001$ (dot-dashed, solid and dotted lines respectively). 
The two dotted lines represent two different string networks in which strings come
inside the horizon at 
the time $t \sim t_{\rm eq}$ and $t \sim 10^{-16} t_{\rm eq}$.}
\label{figpower}
\end{figure}

In figure \ref{figpower} we plot the power spectra of matter perturbations 
for several values of $h$, $\omega_0$ and $\alpha$.
The string mass-per-unit-length $\mu$
in each of the curves has been determined by the CMB normalization \cite{acm}.
In the top panel, the power spectra for $h = 0.65$, $\alpha=0.5$ and
$\Omega_0 = 0.7,\,0.5,\,0.3$ are shown (dot-dashed, solid and dashed lines respectively). 
In the middle pannel we plot the power spectra for $\Omega_0 = 0.5$, $\alpha=0.5$ and 
$h = 0.75,\,0.65,\,0.55$ (dot-dashed, solid and dashed lines respectively). The lower pannel 
represents the power spectra for $\Omega_0=0.5$, $h=0.65$ and 
$\alpha=1,\,0.5,0.001$ (dot-dashed, solid and dotted lines respectively). 
The two dotted lines represent two different string networks in which strings come inside the horizon at 
the time $t \sim t_{\rm eq}$ and $t \sim 10^{-16} t_{\rm eq}$.
In the three descending panels, the individual spectra
are shown with the Peacock and Dodds \cite{pd} 
reconstruction of the linear power spectrum. Since $\Omega_0 < 1$
the reconstructed spectrum has been scaled
as $\propto \Omega^{-0.3}$ (see equation (41) of \cite{pd}).
We can see that hybrid models with $\Gamma=\Omega_0 h \sim 0.3$ and $\alpha\sim0.5$ 
are in reasonable agreement with the observational data. We should point out that although in the 
top two panels active and passive sources contribute equally to the CMB normalization the same does 
not happen to the power spectrum of density perturbations where the passive fluctuations dominate on large 
scales. We can also see that for strings entering the horizon around $t_{eq}$ the amplitude of small 
scale perturbations is clearly reduced with respect to a standard cosmic string model due to absence of 
fluctuations seeded during the radiation era.  As is well known,
this has the desirable 
effect of producing a power spectrum of density fluctuations whose 
shape is in better agreement with observations.

\section{Discussion and conclusions}
\label{soth}

In this paper we have provided a quantitative framework and analysis for Vilenkin's
suggestion \cite{tdi} that cosmic strings can naturally form in open inflation
scenarios. We have used the analytic model of Martins and Shellard \cite{ms2} to
study the evolution of the cosmic string network in this scenario.
As expected, strings are stretched outside the horizon
due to the inflationary expansion but since the number of e-foldings in the second
inflationary stage is relatively small, the ensuing extended epoch of
conformal stretching of the string network network makes it re-enter
the horizon before the epoch of equal matter and radiation densities. We have
determined the power spectrum of cold dark matter perturbations
in these hybrid models, finding good agreement with observations for values of
$\Gamma=\Omega_0h\sim0.3$ and contributions from the active and
passive sources which are comparable in terms of their contributions to the CMB.

We should point out that while this paper was being finalized,
another paper appeared \cite{dterm} discussing another hybrid structure
formation model. This has the string network forming after an epoch of
D-term inflation (thus requiring supersymmetry), and is restricted to
the $\Omega_0=1$ case. Still the results are, as one would expect,
qualitatively very similar. The main difference is that our additional freedom to vary
$\Omega_0$ allows us to obtain slightly better fits to observational data.

Now, given that as we have seen cosmic strings can survive the
second inflationary epoch, one might wonder if the same is true for monopoles.
In other words, how many e-foldings of inflation
do we need in order not to worry about the monopole problem?
The most stringent limit on the abundance of magnetic monopoles formed
in a GUT phase transition is based on the observed luminosity
of the pulsar PSR1929+10. The capture of monopoles
by the pulsar and subsequent energetic bursts due to
monopole catalysis of nucleon decay lead to the flux limit
\begin{equation}
\langle F_M \rangle \le 10^{-27} cm^{-2} sr^{-1} s^{-1}
\end{equation}
(see \cite{kolbetal}).
The predicted flux, however, is
\begin{equation}
\langle F_M \rangle \le 10^{10} \bigl( {n_M \over s} \bigr)
cm^{-2} sr^{-1} s^{-1},
\end{equation}
corresponding to a density
\begin{equation}
\Omega_Mh^2\simeq 10^{24} \left(\frac{n_M}{s}\right)\left(\frac{m_M}{10^{16}{\rm GeV}}\right)
\end{equation}
Thus, we require $n_M/s \ll 10^{-37}$.
The predicted monopole-entropy ratio at reheating, allowing
for $N$ e-foldings of inflation, is
\begin{eqnarray}
{n_M \over s} |_{t_r} &=& {45 \over 2 \pi^2 g_{*s}}
\bigl( \xi_{corr} T_{r} {\rm e}^N \bigr)^{-3} \cr
&=& {45 \over 2 \pi^2 g_{*s}}
\bigl( {T_{pl}\over T_{GUT}} {\rm e}^N \bigr)^{-3} \cr\cr
&\approx& 10^{-12} {\rm e}^{-3 N}.
\end{eqnarray}
This means that $N \gtrsim 20$ e-foldings are enough to satisfy the
monopole problem. We have therefore uncovered a rather interesting cosmological
consequence of relatively short periods of inflation (say 30 to 60
e-foldings)---they will wash away any pre-existing monopoles, but cannot
do so with cosmic strings.

Finally, we emphasize (as was already done by Vilenkin \cite{tdi}) that
the usual bounds on the cosmic string mass per unit length coming from
nucleosynthesis and millisecond pulsar observations do not apply to the open
inflationary scenario presented here. In particular, the fact that strings will
be outside the horizon for most of the radiation era means that there will
be a strong suppression of the `red noise' part of their gravitational wave
spectrum. We hope to discuss these issues in more detail
in future publications.

\acknowledgements

We would like to thank Paul Shellard for useful conversations.
P.P.A. is funded by JNICT (Portugal) under
`Programa PRAXIS XXI' (grant no. PRAXIS XXI/BPD/9901/96).
The work of R.R.C. is supported by the DOE at
Penn (DOE-EY-76-C-02-3071).
C.M. is funded by JNICT (Portugal) under
`Programa PRAXIS XXI' (grant no. PRAXIS XXI/BPD/11769/97).


\vfill\eject

\end{document}